\documentclass[twocolumn,prl,aps]{revtex4-1}
\usepackage{amssymb}
\usepackage{amsmath}
\usepackage{graphicx}
\usepackage{epsfig}

\begin{document}

\title{Visualizing Topology of Real-Energy Gapless Phase Arising from
Exceptional Point}
\author{X. M. Yang}
\author{P. Wang}
\author{L. Jin}
\email{jinliang@nankai.edu.cn}
\author{Z. Song}
\email{songtc@nankai.edu.cn}
\affiliation{School of Physics, Nankai University, Tianjin 300071, China}

\begin{abstract}
The discovery of novel topological phase advances our knowledge of nature
and stimulates the development of applications. In non-Hermitian topological
systems, the topology of band touching exceptional points is very important.
Here we propose a real-energy topological gapless phase arising from
exceptional points in one dimension, which has identical topological
invariants as the topological gapless phase arising from degeneracy points.
We develop a graphic approach to characterize the topological phases, where
the eigenstates of energy bands are mapped to the graphs on a torus. The
topologies of different phases are visualized and distinguishable; and the
topological gapless edge state with amplification appropriate for
topological lasing exists in the nontrivial phase. These results are
elucidated through a non-Hermitian Su-Schrieffer-Heeger ladder. Our findings
open new way for identifying topology phase of matter from visualizing the
eigenstates.
\end{abstract}

\maketitle

\textit{Introduction.---}Topological phase of matter has become a frontier
research field due to their novel features for potential applications \cite%
{Hasan,XLQ,CKC,HMW,LUL,Khanikaev,Goldman,Cooper,OzawaT,Hafezi,Rechtsman,WJChen,Mukherjee,Bandres,YDChong16,Lin,JWDong,Rechtsman17,Mittal,Klembt}%
. Majorana \cite{LF,RML,VM,YO}, Dirac \cite{AHCN,ZKL,JAS,ZW,JX,SMY}, and
Weyl fermions \cite{MH,SMH,BQL,SYX,Vishwanath} predicted in high-energy
physics are discovered in this fertile ground. These concepts stimulate an
interesting topic of topological gapless phases/semimetals \cite%
{Yang,Burkov,Xu,Kim,Wang1,Hou1,Sama,Neupane,Lu}. The symmetry protected
nodal points in topological gapless phase are topological defects of an
auxiliary vector field and are irremovable until they meets and annihilates
in pairs. The topology of gapless bands can be characterized by kink in
one-dimension (1D) \cite{Li} or vortex of skyrmion in two-dimension (2D)
\cite{Hou1,Sama}.

Nowadays, the great efforts have been made to unravel the mystery of
non-Hermitian physics \cite%
{Bender,Ali1,NMBook,Ruschhaupt,Klaiman,Makris08,Guo,Ruter,Peng,XZhang,Chang,Fleury,Makris15,LFengReview,El,Gupta}%
. Non-Hermitian phase transition occurs at exceptional point (EP) \cite%
{Guo,Ruter,Peng,Chang}, which is a unique concept in non-Hermitian systems
and has exotic topology related to Riemann surface \cite%
{Dembowski,EP,Uzdin,Zhen,Doppler,Xu1,ZPLiu,Wiersig,Chen,Hodaei,QZhong,KDing,KDingPRL,Zhang,Midya,Miri}%
. The investigations of topological physics have been extended to
non-Hermitian region \cite%
{Levitov,Diehl,YFChen,Zeuner,HZhao,GQLiang,SChen,2011,Schomerus,JGong15,Song,YXu2,HShen2,WZhu,Yuce,Lieu,Hirschmann,Bergholtz,KunstTM,KK,JHou,Takata,CYZhao,Pump,VMMA,Herviou,Yoko,Schnyder,DasReview,TOhashi}%
. The non-Hermitian topological systems under various combined symmetries
including the parity-time ($\mathcal{PT}$) symmetry are investigated \cite%
{Malzard2,JL,LS,PXue,XNi,LJLPRB,Koutserimpas,Cancellieri,AYoshida}. The
non-Hermitian topological invariant and band theory are established \cite%
{Xu2,Leykam,Shen}. The $\mathcal{PT}$-symmetric interface states are
realized in passive optical systems \cite{Poli,Weimann,MPan}. Furthermore,
the edge state lasing is demonstrated in active optical systems \cite%
{Jean,Parto,Feng,Harari,Kartashov,Secli}. The high-order non-Hermitian
topological systems \cite{TLiu,EzawaPRB,CHLeeHighOrder,XWLuo,Edvardsson},
the breakdown of bulk-boundary correspondence \cite%
{TELee,Torres,ZWang,LJinPRB,CHLee,JHu,HZhang,1819,ZGong}, and the
topological classifications \cite{ZGong,Kawabata1,Kawabata2,KawabataEP,HZhou}
are studied.

In non-Hermitian topological systems, the gap closing in energy band is
usually associated with EPs instead of degenerate points (DPs) \cite%
{Xu2,Leykam,Shen,Zhou2,Szameit}. The band touching EP pairs split from
single DPs are connected by open Fermi arcs \cite%
{Zhou2,SLin,Kozii,Schomerus18}; alternatively, non-Hermitian semimetals
exhibit nodal phases with symmetry protected EPs rings and surfaces \cite%
{ZhouH,Cerjan1,Cerjan2,Okugawa,Budich,Yoshida,RAMolina}; the corresponding
energy bands in the non-Hermitian gapless phases are all complex. The
complex-energy bands associated EPs are systematically studied \cite%
{KawabataEP} and they are dramatically different from the gapless phase in a
Hermitian system. In contrast to the Hermitian topological gapless phase, a
non-Hermitian topological gapless phase is typically characterized by two
types of winding numbers; an additional winding number solely for
non-Hermitian systems is defined to characterize the topology of Riemann
sheet energy bands \cite{Leykam,Shen,ZGong}. The real-energy energy band
does not exponentially decay or increase as time and has a zero winding in
contrast to the complex energy band. Thus, two winding numbers are
insufficient to distinguish the gapless phases arising from EPs and DPs.

The aim of this Letter is to propose a real-energy gapless phase associated
with band touching EPs. The balanced gain and loss are introduced in chiral
symmetric Hermitian topological insulators to create the gapless phase
arising from EPs, where robust zero modes with amplification and attenuation
are generated for nontrivial topology. We offer a powerful graphic approach
to visualize the topological features of different phases. The rich
topological phases of either gapped (separable) or gapless, either Hermitian
or non-Hermitian band touching, and either trivial or nontrivial phases are
all distinguishable from the geometrical topologies of eigenstate graphs.
The essence of eigenstate graphs dramatically differs from the knotted or
linked nodal lines in semimetals that representing the zero-energy
(equal-energy) surface \cite{ZhouH,Cerjan1,Carlstrom}. In the graphic
approach, the graphic eigenstates of real gapless bands arising from EPs
form a network; the topological nature of which is characterized by its
nodes, branches, and independent loops. The networks with different links,
fixed/movable nodes correspond to different topological phases; other
gapless and gapped phases possess distinct geometric graphs.

\textit{Gapless phase arising from EPs.---}In the theory for topological
insulators \cite{Hasan,XLQ}, the energy band is insufficient to determine
the full topological character of a phase of matter. The bulk topological
features are encoded in the eigenstates. We consider a prototype of
non-Hermitian topological system: the 1D two-band model. In the momentum
space, the core matrix reads $h_{k}=\mathbf{B}\cdot \mathbf{\sigma }$, where
$\mathbf{B}=\left( B_{x},B_{y},B_{z}\right) $ is an effective magnetic
field, and $\mathbf{\sigma }=\left( \sigma _{x},\sigma _{y},\sigma
_{z}\right) $ is the Pauli matrix; $B_{x}$ and $B_{y}$ are real, but $B_{z}$
is imaginary. The energy bands are tighten under the influence of $B_{z}$;
however, the topological phase transition in $h_{k}$ is not affected by $%
B_{z}$. Thus, the gapless phase is created via increasing $B_{z}$ in the
gapped phase till a non-Hermitian phase transition.

\textit{Graphic eigenstate.---}For $B^{2}=B_{x}^{2}+B_{y}^{2}+B_{z}^{2}%
\geqslant 0$, the eigen values $\varepsilon _{k}^{\pm }=\pm B$ are real with
corresponding eigenstates%
\begin{equation}
\left\vert \psi _{k}^{\pm }\right\rangle =\frac{1}{\sqrt{2}}\left(
\begin{array}{c}
e^{i\varphi _{\pm }(k)} \\
1%
\end{array}%
\right) ,
\end{equation}%
where $\varphi _{+}(k)=\arctan \left( -B_{y}/B_{x}\right) +\arctan \left(
-iB_{z}/B\right) $\ and $\varphi _{-}(k)=\arctan \left( -B_{y}/B_{x}\right)
\pm \pi -\arctan \left( -iB_{z}/B\right) $ are real functions with a period $%
2\pi $. The varying directions of $\varphi _{\pm }(k)$ and $\mathbf{B}$ are
the same. Therefore, $\varphi _{\pm }(k)$ as the kernel of eigenstates $%
\left\vert \psi _{k}^{\pm }\right\rangle $ provides the information of the
topological nature of $h_{k}$. Notably, $\left\vert \psi _{k}^{\pm
}\right\rangle $ can be represented by a loop on the torus spanned by $%
\varphi _{\pm }(k)$ and $k$, referred to as the graphic eigenstates.

In the $k$ space, $h_{k}$ is $\mathcal{PT}$-symmetric $\left( \mathcal{PT}%
\right) ^{-1}h_{k}\left( \mathcal{PT}\right) =h_{k}$ and $\mathcal{CT}$%
-symmetric $\left( \mathcal{CT}\right) ^{-1}h_{k}\left( \mathcal{CT}\right)
=-h_{-k}$, where the operators $\mathcal{PT=\sigma }_{x}\mathcal{K}$, $%
\mathcal{CT}=\sigma _{z}\mathcal{K}$, and $\mathcal{K}$ is the complex
conjugation. The $\mathcal{PT}$ symmetry is related to the reality of the
spectrum; while the $\mathcal{CT}$ symmetry protects the system topology.
For an arbitrary eigenstate $\left\vert \psi _{k}^{\pm }\right\rangle $, the
$\mathcal{CT}$ symmetry requires another eigenstate satisfying $\left\vert
\psi _{-k}^{\mp }\right\rangle =\mathcal{CT}\left\vert \psi _{k}^{\pm
}\right\rangle $, which leads to%
\begin{equation}
\varphi _{+}\left( k\right) +\varphi _{-}\left( -k\right) =\pm \pi .
\end{equation}%
In particular, when the EP appears at $k_{c}$, two eigenstates $\left\vert
\psi _{k_{c}}^{+}\right\rangle $ and $\left\vert \psi
_{k_{c}}^{-}\right\rangle $ coalesce to one denoted as $\left\vert \psi
_{k_{c}}^{0}\right\rangle $. The $\mathcal{CT}$ symmetry ensures $\left\vert
\psi _{-k_{c}}^{0}\right\rangle =\mathcal{CT}\left\vert \psi
_{k_{c}}^{0}\right\rangle $, i.e., the existence of a pair of EPs with zero
energy for $k_{c}\neq 0,\pi $; any EP cannot be separately removed, but the
EP position changes as the system parameters. When $k_{c}=0,\pi $, we have
one fixed EP with $\varphi _{\pm }(k_{c})=\pi /2 $ or $-\pi/2$.

For a Hermitian system with $B_{z}=0$, we always have $\varphi _{-}\left(
k\right) =\varphi _{+}\left( k\right) \pm \pi $. In the presence of a DP,
two energy bands usually form a single knot [Fig.~\ref{fig1}(c)]. In the
absence of DPs, $\mathbf{B\neq 0}$, two energy bands form two loops without
intersection; this is similar as the gapped non-Hermitian phase shown in
Figs.~\ref{fig1}(a) and \ref{fig1}(b). For a non-Hermitian system, the
effective magnetic field $\mathbf{B}$ is complex. In contrast to two
eigenstates of a Hermitian matrix that represented by two opposite points on
the torus due to the orthogonality, two eigenstates of the real-energy bands
in a non-Hermitian matrix represented by two points can have arbitrary
positions on the torus; and they meet when eigenstates coalesce at the EPs,
where $\varphi _{-}\left( k\right) =\varphi _{+}\left( k\right) $. Then, two
graphic eigenstates constitute a network [Figs.~\ref{fig1}(e)-1(h)]. The
network on the $\varphi $-$k$ torus provides a complete topological picture
for the real-energy gapless phase arising from EPs, the topological features
of which are reflected from the network topology. Notably, the graphs
represent the eigenstate rather than the zero energy, which essentially
differs from the EP links or knotted nodal lines of EPs \cite%
{JHu,HZhang,Budich,Okugawa,Cerjan1,ZhouH,Cerjan2,Yoshida,RAMolina,Carlstrom}.

\begin{figure}[tbp]
\includegraphics[ bb=0 0 950 605, width=8.8 cm, clip]{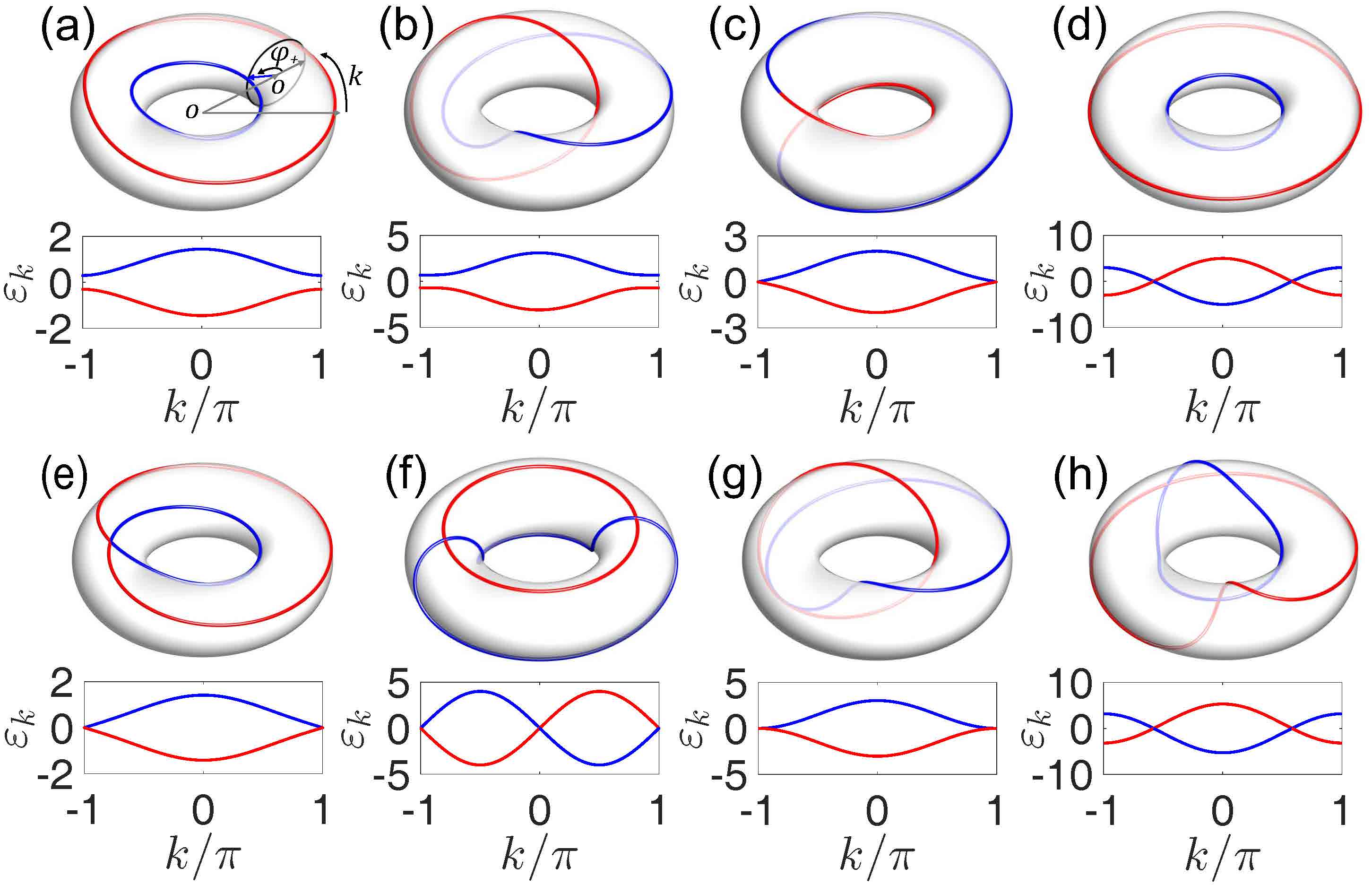}
\caption{Graphic eigenstates and energy bands of the SSH ladder. $t=1$ and $(v,w,\protect\gamma )$ is (a) $(1/4,1/4,2/5)$, (b) $(2,2/7,1)$, (c) $(1/4,3/4,0)$,
(d) $(2,2,0)$, (e) $(1/4,1/4,\protect\gamma _{c})$, (f) $(-2,2,\protect\gamma _{c})$, (g) $(2,2/7,\protect\gamma _{c})$, and (h) $(3,3/2,\protect\gamma _{c})$. $\protect\gamma _{c}$ is chosen Eq.~(\protect\ref{gamma2}) in (e) and (f); but chosen Eq.~(\protect\ref{gamma1}) in (g) and (h). (c) and (d) are the Hermitian case; in the non-Hermitian case, (a), (e), and (f) [(b), (g), and (h)] are in the topologically trivial (nontrivial) phase. The gapless phases arising from EPs are (e)-(h).}
\label{fig1}
\end{figure}

\textit{Topological characterization.---}In general cases, multiple EPs may
appear; accordingly, the graphic eigenstates may form complicate networks.
The geometrical topology of any network satisfies a fundamental fact: the
numbers of nodes $n$, independent loops $l$, and branches $b$ fulfill%
\begin{equation}
n=b-l+1.
\end{equation}%
An independent loop must contain at least one branch that not belongs to any
other loop. For instance, we have $n=1\left( 2\right) ,b=2\left( 4\right)
,l=2\left( 3\right) $ for the network shown in Fig.~\ref{fig1}(g) [Fig.~\ref%
{fig1}(h)]. This graphic approach is especially helpful in characterizing
the real-energy topological gapless phase arising from EPs.

Two graphic eigenstate loops are unlinked [Fig.~\ref{fig1}(a)] or linked
[Fig.~\ref{fig1}(b)], which indicate the trivial or nontrivial topology. To
verify this assertion, we point out that the varying direction of $\varphi
_{\pm }\left( k\right) $ accumulated in a period of $k$ defines a winding
number%
\begin{equation}
\mathcal{N}_{\pm }=\left( 2\pi \right) ^{-1}\int_{0}^{2\pi }\nabla
_{k}\varphi _{\pm }(k)\mathrm{d}k.
\end{equation}

Besides, the average values of Pauli matrices under two eigenstates $|\psi
_{k}^{\pm }\rangle $ yield $\left\langle \sigma _{x}\right\rangle _{\pm
}=\cos [-\varphi _{\pm }(k)]$, $\left\langle \sigma _{y}\right\rangle _{\pm
}=\sin [-\varphi _{\pm }(k)]$, and $\left\langle \sigma _{z}\right\rangle
_{\pm }=0$ \cite{SI}; which define a planar vector field that associated
with the upper ($\left\vert \psi _{k}^{+}\right\rangle $) or lower ($%
\left\vert \psi _{k}^{-}\right\rangle $) band%
\begin{equation}
\mathbf{F}_{\pm }\left( k\right) =(\left\langle \sigma _{x}\right\rangle
_{\pm },\left\langle \sigma _{y}\right\rangle _{\pm }).
\end{equation}%
$\mathbf{F}_{\pm }\left( k\right) $ characterizes the topological properties
of energy band \cite{Hou1,Sama, LJinPRB,SLin}. The winding number associated
with $\mathbf{F}_{\pm }\left( k\right) $ is $\left( 2\pi \right)
^{-1}\int_{0}^{2\pi }\nabla _{k}\arg \mathbf{F}_{\pm }\left( k\right)
\mathrm{d}k$, which equals to the winding number $\mathcal{N}_{\pm }$. The
coincidence of varying directions of $\varphi _{\pm }(k)$ and $\mathbf{F}%
_{\pm }(k)$ indicates that the topological properties of energy band are
reflected from the kernel of eigenstate $\varphi _{\pm }(k)$. $2\pi \mathcal{%
N}_{\pm }$ is the nonzero rotation angle of $\varphi _{\pm }(k)$ accumulated
in a period of $k$, indicating the nontrivial topology of the gapped or
gapless phase, and the graphic eigenstate loops are linked. Notably, the
winding numbers for the two bands are identical $\mathcal{N\equiv N}_{+}=%
\mathcal{N}_{-}.$

For the gapless phase with EPs, the topology of EPs arising from entirely
real-energy bands are in a striking difference from the EPs in the complex
energy bands. The difference is revealed from another winding number $%
\mathcal{W}_{\mathrm{EP}}=\left( 2\pi \right) ^{-1}\int_{-\pi }^{\pi }\nabla
_{k}\arg E_{+}\left( k\right) \mathrm{d}k$ that characterizing the Riemann
sheets \cite{Leykam,Shen,ZGong}. Notably, $\mathcal{W}_{\mathrm{EP(DP)}}=0$
for the EPs (DPs) of real-energy gapless bands in contrast to $\mathcal{W}_{%
\mathrm{EP}}=\mathcal{\pm }1/2$ for the EPs in the complex energy bands \cite%
{Leykam,Zhou2,KK}. However, the topologies of real-energy gapless phases
arising from DPs and EPs dramatically differ from each other; the node
(network configuration) is absent for the gapless bands with DPs, while the
presence of node is a typical feature for the gapless bands with EPs.

\textit{Non-Hermitian SSH ladder.---}As a prototype of one-dimensional
non-Hermitian topological system, the $\mathcal{PT}$-symmetric non-Hermitian
Su-Schrieffer-Heeger (SSH) model is experimentally realized in the coupled
waveguides \cite{Zeuner,MPan,Weimann}, coupled resonators \cite{Poli}, and
polariton micropillars \cite{Jean}. The essential features of the $\mathcal{%
PT}$-symmetric non-Hermitian system are captured even though the system is
passive; the active non-Hermitian SSH model is realized with additional
pumping \cite{Jean,Feng,Parto}. Candidates for the experimental realization
of non-Hermitian topological systems include photonic crystals \cite%
{LUL,Leykam16,OzawaT}, ultracold atomic gasses \cite{Cooper,Diehl}, acoustic
lattices \cite{HeC,Xiao,Fleury2}, and electric circuits \cite%
{Ezawa,EzawaPRB,ChongNC,Lee}. We employ a non-Hermitian SSH ladder [Fig.~\ref%
{fig2}(a)] to elucidate our findings. Introducing additional loss $-2i\gamma
$ in one sublattice (the pink lattice) generates a passive non-Hermitian SSH
ladder \cite{Guo}; an overall decay rate $-i\gamma $\ offset in both
sublattices yields a $\mathcal{PT}$-symmetric non-Hermitian SSH ladder \cite%
{Malzard2}; alternatively, we can consider an active $\mathcal{PT}$%
-symmetric non-Hermitian SSH ladder constituted by two coupled SSH chains
incorporated gain and loss \cite{Parto}. The Hamiltonian in the real space
reads%
\begin{equation}
H=\sum_{j=1}^{N}(wa_{j+1}^{\dagger }b_{j}+va_{j}^{\dagger
}b_{j+1}+ta_{j}^{\dagger }b_{j}+\mathrm{H.c.})+i\gamma (a_{j}^{\dagger
}a_{j}-b_{j}^{\dagger }b_{j}),  \label{SSH chain}
\end{equation}%
where $a_{j}^{\dag }$ ($b_{j}^{\dag }$) is the creation operator of the $j$%
th site in the sublattice $A$ ($B$). $H$ is a two-leg ladder with $2N$
sites, each leg is a $\mathcal{PT}$-symmetric SSH chain with staggered real
couplings $w$ and $v$ \cite{Poli,Weimann,MPan,Jean,Feng,Parto}; two ladder
legs are coupled at the strength $t$ after one leg glided by one site. $H$
is a simple generalization of the $\mathcal{PT}$-symmetric non-Hermitian SSH
model. Applying the Fourier transformation, the Hamiltonian under periodical
boundary condition is rewritten as $H=\sum_{k}h_{k}=\sum_{k}\mathbf{B}\cdot
\mathbf{\sigma }$. The core matrix reads%
\begin{equation}
h_{k}=\left(
\begin{array}{cc}
i\gamma & we^{ik}+ve^{-ik}+t \\
we^{-ik}+ve^{ik}+t & -i\gamma%
\end{array}%
\right) .  \label{hk}
\end{equation}%
The non-Hermitian SSH ladder has $\mathcal{PT}$ symmetry and chiral-time ($%
\mathcal{CT}$) symmetry \cite{Schomerus,Malzard2,LS}. The corresponding
energy bands $\varepsilon _{k}^{\pm }$ for the graphic eigenstates are
depicted as a function of $k$ in the lower panel of Fig.~\ref{fig1}.

\begin{figure}[tbp]
\includegraphics[ bb=5 5 515 345, width=8.8 cm, clip]{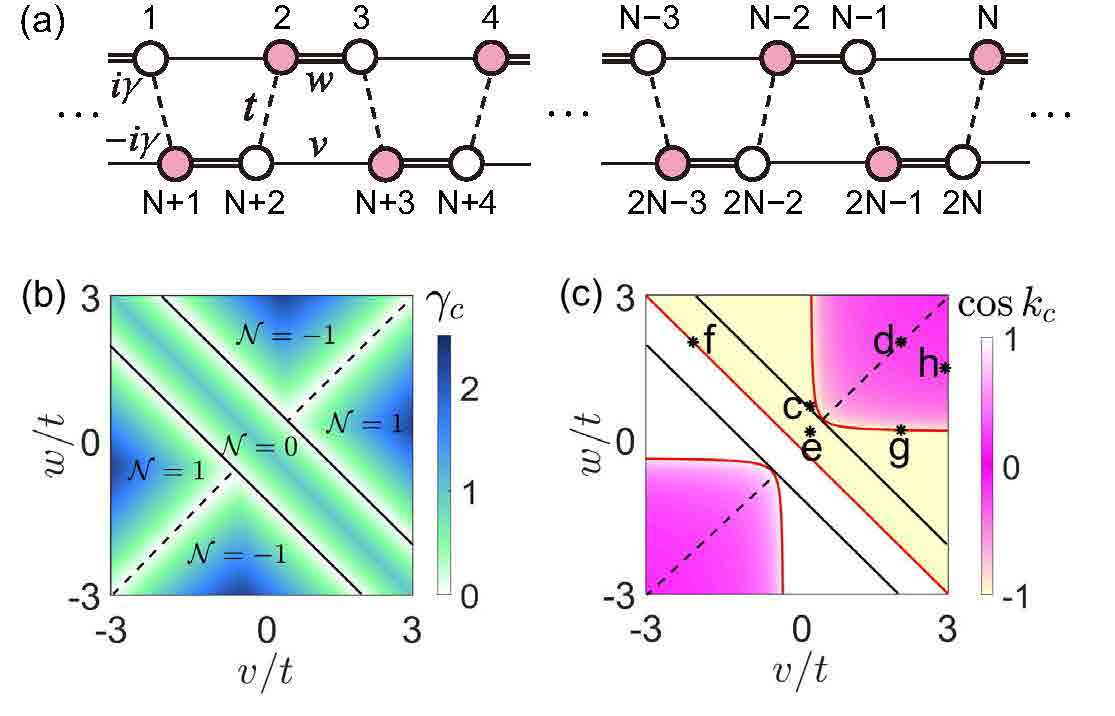}
\caption{(a) Schematic of the non-Hermitian SSH ladder. The gain (loss) is in sublattice $A$ ($B$) in white
(pink). (b) Phase diagram and critical $\protect\gamma _{c}$ [Eqs.~(\protect\ref{gamma1}) and (\protect\ref{gamma2})]. (c) $\cos(k_c)$ for the EPs. Red solid curves divide the $v$-$w$ plane into four regions; the band gap closes at
EPs at $k_{c}$, $0$ and $\protect\pi $. Two hyperbola
curves are $\left\vert w+v\right\vert =4wv/t.$ Stars $c$ to $f$ are marked for the cases in Fig.~\ref{fig1}.}
\label{fig2}
\end{figure}

In the gapped phase, the graphic eigenstates are two separated loops on the $%
\varphi $-$k$ torus. The two loops are unlinked [Fig.~\ref{fig1}(a)] in the
topologically trivial phase with $\mathcal{N}=0$; and are linked [Fig.~\ref%
{fig1}(b)] in the topologically nontrivial phase with $\mathcal{N}=\pm 1$.
The system has the gapless phase arising from DPs in the Hermitian case; at $%
w+v=\pm t$, we notice one fixed DP [black solid line in Fig.~\ref{fig2}(b)],
and two movable DPs at $w=v$ [black dashed line in Fig.~\ref{fig2}(b)] in
region $\left\vert w+v\right\vert >t$ that characterized by the vector field
kinks \cite{Li}. One fixed DP appears at $k_{c}=0$ or $\pi $. The energy
bands constitute a single band and the graphic eigenstates form a knot [Fig.~%
\ref{fig1}(c)]. At $w=v$, two components of the magnetic field $\mathbf{B}$
vanish, the band gap closes and two movable DPs appear at $\cos
k_{c}=-t/\left( 2v\right) $. Two loops on the $\varphi $-$k$ torus are
separated without any intersection [Fig.~\ref{fig1}(d)].

As gain and loss increase, the band gap shrinks and closes at the critical
non-Hermiticity%
\begin{equation}
\gamma _{c}=\left\vert v-w\right\vert \sqrt{1-t^{2}/\left( 4vw\right) },
\label{gamma1}
\end{equation}%
in the region $\left\vert w+v\right\vert \leqslant 4wv/t$, where the EPs are
movable in the momentum space and appear at $\cos k_{c}=-t\left( w+v\right)
/\left( 4wv\right) $; otherwise, the band gap closes at the critical
non-Hermiticity%
\begin{equation}
\gamma _{c}=\left\vert w+v\pm t\right\vert ,  \label{gamma2}
\end{equation}%
and the EPs are fixed in the momentum space and appear at $k_{c}=0$ or $\pi $%
. $\gamma _{c}$ for the band gap closing is depicted in Fig.~\ref{fig2}(b); $%
\cos k_{c}$ for the location of EPs are depicted in Fig.~\ref{fig2}(c). In
Fig.~\ref{fig2}(b), the black dashed line separates two phases with winding
numbers $\mathcal{N}=1$ and $\mathcal{N}=-1$; while the black solid line
separates phases with winding numbers $\mathcal{N}=\pm 1$\ and $\mathcal{N}%
=0 $.

The hyperbola $\left\vert w+v\right\vert =4wv/t$\ and line $w+v=0$ divide
the $w$-$v$ plane into four real-energy gapless phases with distinct
topological characters [Fig.~\ref{fig2}(c)]. The topological feature is
clearly revealed by the graphic approach: (i) For $\gamma <\gamma _{c}$, two
real-energy bands are separated without any EP; two loops on the $\varphi $-$%
k$\ torus have none intersection [Figs.~\ref{fig1}(a) and (b)]. (ii) For $%
\gamma =\gamma _{c}$\ in the region $\left\vert w+v\right\vert <4wv/t$\
except for $w=v$, two loops on the $\varphi $-$k$ torus have two robust
nodes [Fig.~\ref{fig1}(h)], which are movable but irremovable as the system
parameters $w$ and $v$; two movable nodes merge to a single fixed node at $%
k_{c}=0$ or $\pi $ associated with the change of network topology at
hyperbola $\left\vert w+v\right\vert =4wv/t$ [Fig.~\ref{fig1}(g)]. (iii) For
$\gamma =\gamma _{c}$ in the region $\left\vert w+v\right\vert \geqslant
4wv/t$, two loops have one fixed node except for $w+v=\pm 1$, the location
of which does not change as the parameters $w$ and $v$ [Fig.~\ref{fig1}(e)].
(iv) For $\gamma =\gamma _{c}$\ and $\left\vert w+v\right\vert =0$, two
loops have two fixed nodes [Fig.~\ref{fig1}(f)]. The graphic eigenstates are
depicted for $\gamma >\gamma _{c}$ in Supplemental Material \cite{SI}, where
the loops detach the $\varphi $-$k$ torus in the broken $\mathcal{PT}$%
-symmetric phase.

\begin{figure}[tbp]
\includegraphics[ bb=15 60 870 555, width=8.8 cm, clip]{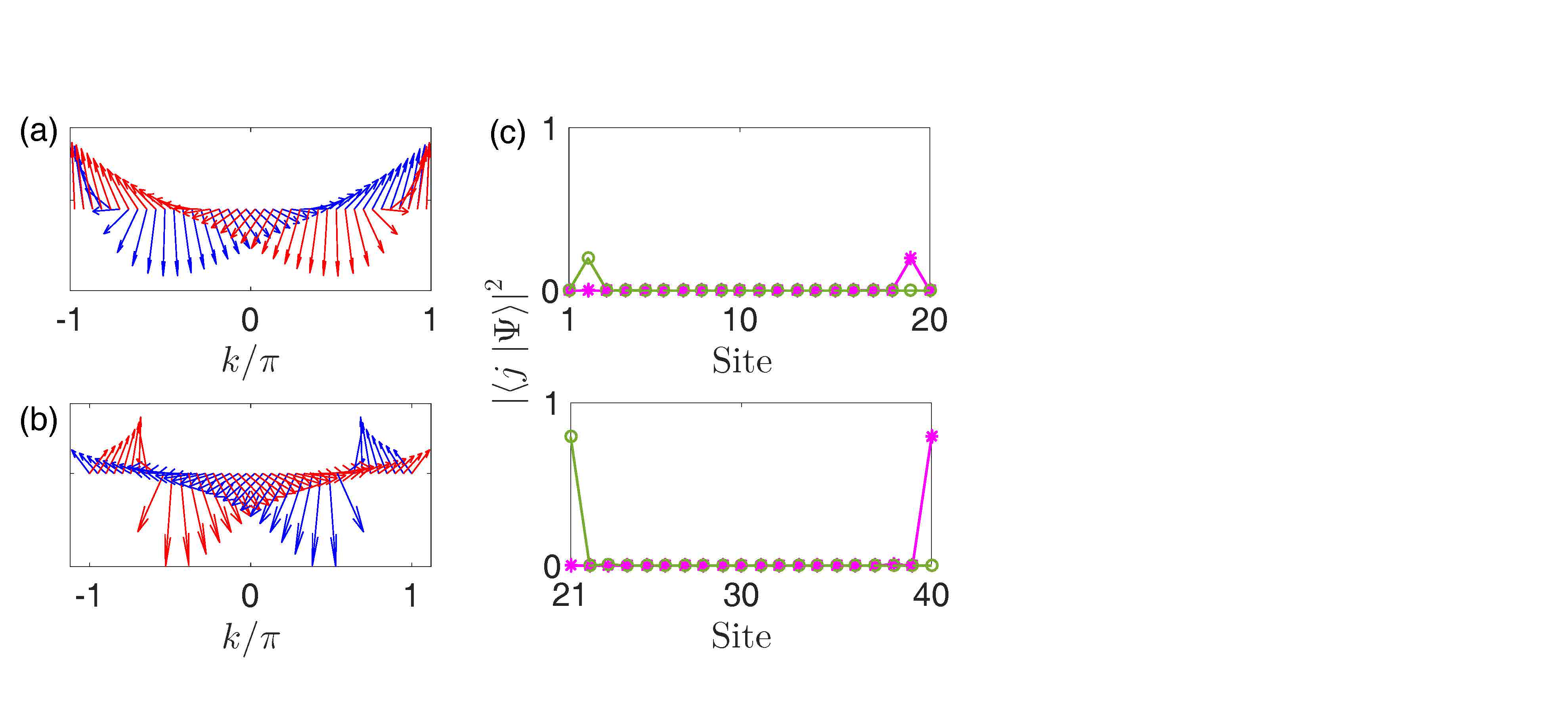}
\caption{Vector field $\mathbf{F}_{+}\left( k\right)$ [$\mathbf{F}_{-}\left(
k\right) $] for the upper (lower) energy band represented by the blue (red)
arrow at (a) $(v,w,\protect\gamma)=(2, 2/7, \gamma _{c})$, (b)
$(v,w,\protect\gamma)=(3, 3/2, \gamma _{c})$; other parameter is $t=1$. (c)
Gapless zero modes with amplification ($E_{\mathrm{ZM}}=i\protect\gamma $)
and attenuation ($E_{\mathrm{ZM}}=-i\gamma $) for (a), $N=20$. The zero mode with
amplification locates on the right. The upper and lower panels in (c) are for the upper and lower ladder legs in Fig.~\ref{fig2}(a), respectively.}
\label{fig3}
\end{figure}

The two loops are unlinked in Figs.~\ref{fig1}(e) and \ref{fig1}(f); while
linked in Figs.~\ref{fig1}(g) and \ref{fig1}(h) due to the $2\pi $ variation
of $\varphi _{\pm }(k)$ for each eigenstate in a $2\pi $ period of $k$. This
indicates the topologically nontrivial features and coincides with the
winding number $\mathcal{N}$ marked in Fig.~\ref{fig2}(b). Both two states
give identical rotating angle $2\pi $ after $k$ varying a period. Figures~%
\ref{fig3}(a) and \ref{fig3}(b) depict the vector field $\mathbf{F}_{\pm
}=(\left\langle \sigma _{x}\right\rangle _{\pm },\left\langle \sigma
_{y}\right\rangle _{\pm })$. The $2\pi $ varying direction of the vector
field yields the nontrivial topology of eigenstates and coincides with the
graphic eigenstates shown in Figs.~\ref{fig1}(g) and~\ref{fig1}(h). The
coincidence of the blue and red arrows implies the locations of EPs. The
upper panel depicts the situation with single EP located at $k_{c}=\pi $;
the lower panel depicts the situation with two EPs located at $\cos
k_{c}=-1/4$.

The nontrivial winding of eigenstate predicts the appearance of edge states.
In topologically nontrivial gapless phase, the bands touch at zero energy
and a pair of zero modes appear in the absence of gain and loss; under the $%
\mathcal{CT}$ symmetry, the gapless zero modes become imaginary in the
presence of gain and loss. As the gain and loss are respectively introduced
in two sublattices, thus the left side and right side localized edge states
confined in one sublattice experiences either an additional attenuation $%
-i\gamma $ or an additional amplification $i\gamma $. The probability
distribution of edge states is depicted in Fig.~\ref{fig3}(c) for the
situation in Fig.~\ref{fig3}(a). The amplified edge state is appropriate for
robust topological lasing.

\textit{Conclusion.}---Real-energy topological gapless phase arising from
EPs in 1D is proposed and investigated through a graphic approach. The
graphic eigenstates completely encode the information of system topology and
visualize the topologies of different phases in both Hermitian and
non-Hermitian systems. The graph of real-energy topological gapless phase
arising from EPs forms a network, its geometric topology reflects the
topological properties of the ground state phase diagram. The chiral-time
symmetry protects the topological gapless phase. A pair of amplification and
attenuation topologically zero modes exist in the nontrivial phase. These
are elucidated through a non-Hermitian SSH ladder with balanced gain and
loss that directly accessible in experiment in many physical systems. Our
findings propose a novel non-Hermitian topological gapless phase and provide
insights into topological characterization of topological phase of matter.

\acknowledgments This work was supported by National Natural Science
Foundation of China (Grants No.~11874225 and No.~11605094).

\clearpage

\newpage
\begin{widetext}
\section{Supplemental Material for ``Visualizing Topology of Real-Energy
Gapless Phase Arising from Exceptional Point"}
\begin{center}
X. M. Yang, P. Wang, L. Jin, and Z. Song\\[2pt]
\textit{School of Physics, Nankai University, Tianjin 300071, China}
\end{center}
\subsection{A: Vector field}

The core matrix in the momentum space reads $h_{k}=\mathbf{B}\cdot \mathbf{%
\sigma =}B_{x}\sigma _{x}+B_{y}\sigma _{y}+B_{z}\sigma _{z}$. We consider $%
B_{x}$ and $B_{y}$ are real, but $B_{z}$ is imaginary, then the eigenstates
of $h_{k}$ within the range of real eigenvalues $\varepsilon _{k}^{\pm }$
can be written in the form of Eq. (1) of the Letter, and the average values of Pauli
matrices can be expressed as%
\begin{equation}
\begin{array}{l}
\left\langle \sigma _{x}\right\rangle _{\pm }=\left\langle \psi _{k}^{\pm
}\right\vert \sigma _{x}\left\vert \psi _{k}^{\pm }\right\rangle =\cos
\left( -\varphi _{\pm }\right) \\
\left\langle \sigma _{y}\right\rangle _{\pm }=\left\langle \psi _{k}^{\pm
}\right\vert \sigma _{y}\left\vert \psi _{k}^{\pm }\right\rangle =\sin
\left( -\varphi _{\pm }\right) \\
\left\langle \sigma _{z}\right\rangle _{\pm }=\left\langle \psi _{k}^{\pm
}\right\vert \sigma _{z}\left\vert \psi _{k}^{\pm }\right\rangle =0%
\end{array}%
.
\end{equation}

The average values of Pauli matrices define a vector field that related to
the topological features of the non-Hermitian topological system. The vector
field $\mathbf{F}_{\pm }\left( k\right) =(\left\langle \sigma
_{x}\right\rangle _{\pm },\left\langle \sigma _{y}\right\rangle _{\pm })$
directly relates to the phase factor $\varphi _{\pm }$ of the eigenstates.
For the non-Hermitian SSH ladder, the core matrix is in Eq.~(6) of the
Letter. Within the range of real eigenvalues $\varepsilon _{k}^{\pm },$ the
corresponding eigenstates of $h_{k}$ in Eq.~(1) of the Letter have $\varphi
_{+}=\theta _{1}+\theta _{2}$\thinspace\ and $\varphi _{-}=\theta
_{1}-\theta _{2}\pm \pi $, and $\tan \theta _{1}=\left[ \left( w-v\right)
\sin k\right] /\left[ \left( v+w\right) \cos k\right] $, $\tan \theta
_{2}=\gamma /(|we^{-ik}+ve^{ik}+t|^{2}-\gamma ^{2})^{1/2}$. The complete
information of graphic eigenstates are mapped to the phases $\varphi _{\pm
}\left( k\right) $.

\subsection{B: Graphic eigenstates in the broken phase}

Under the periodical boundary condition of the non-Hermitian SSH ladder, the
effective complex magnetic field is $B_{x}=\left( v+w\right) \cos k+t$, $%
B_{y}=\left( v-w\right) \sin k$, and $B_{z}=i\gamma $ in the core matrix $%
h_{k}$. For large non-Hermiticity $\gamma >\gamma _{c}$ in the broken
parity-time phase, the complex energy levels appear; and the corresponding
eigenstates of $h_{k}$ reduce to the form of%
\begin{equation}
\left\vert \psi _{k}^{\pm }\right\rangle =\left(
\begin{array}{c}
\sin (\theta _{\pm }/2)e^{i\varphi _{\pm }} \\
\cos (\theta _{\pm }/2)%
\end{array}%
\right) .  \label{Psi}
\end{equation}

\begin{figure}[h]
\includegraphics[ bb=0 0 953 368, width=17.8 cm,clip]{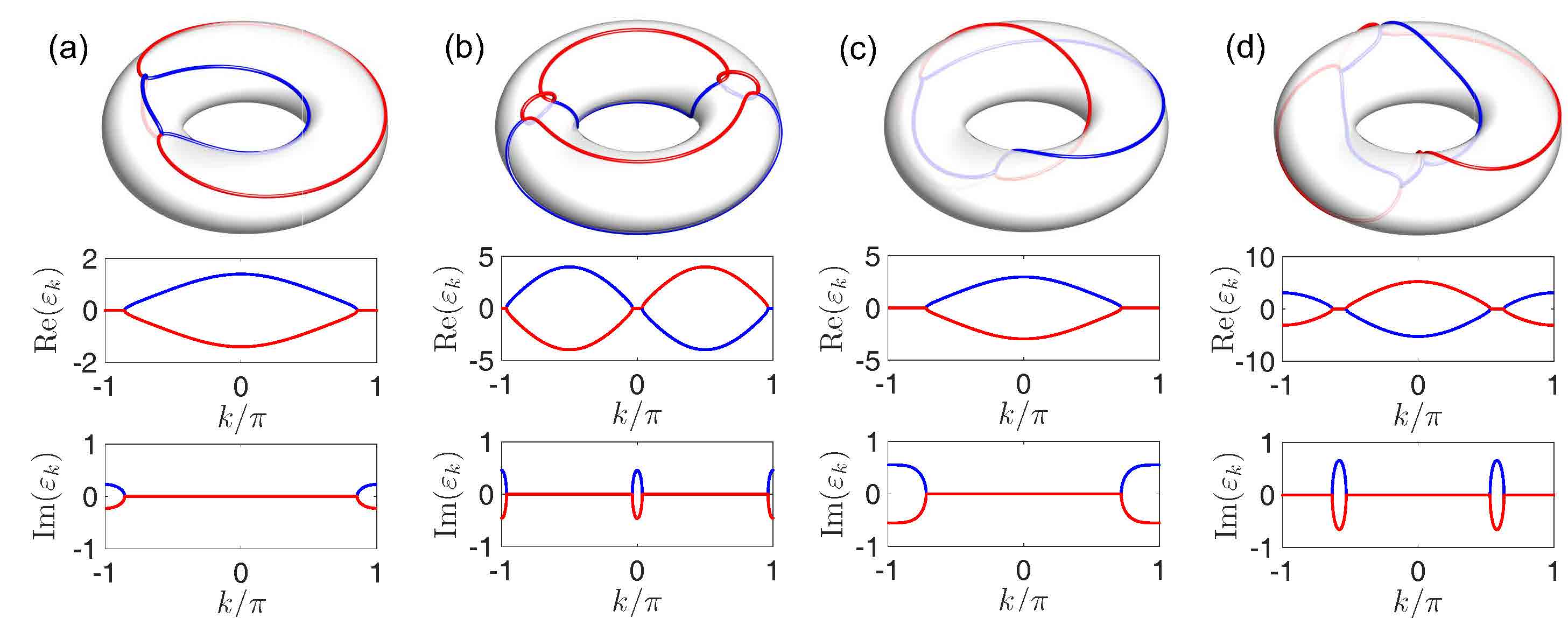}
\caption{Graphic eigenstates and energy bands in the broken $\mathcal{PT}$-symmetric phase
of the non-Hermitian SSH ladder. $(v,w,\protect\gamma)$ is (a) $\left(
1/4,1/4,11/20\right) $, (b) $\left(-2,2,11/10\right) $, (c) $\left( 2,2/7,7/5\right) $ , (d) $\left( 3,3/2,8/5\right)$. In all plots, $t=1$.} %
\label{fig1}
\end{figure}

For the complex $\varepsilon _{k}$, $\varphi _{+}=\varphi _{-}=\arctan
\left( -B_{y}/B_{x}\right) $ and \ $\theta _{\pm }=2\arccos (1/\sqrt{C_{\pm }%
})$ where $C_{\pm }=\left\vert B_{z}\pm B\right\vert ^{2}/\left(
B_{x}^{2}+B_{y}^{2}\right) +1$ is the normalization coefficient. In the $%
\varphi $-$k$ torus, $R_{0}$ is the distance from the center of the tube to
the center of the torus, and $r_{0}$ is the radius of the tube. Two loops
are plotted on the torus with $R=R_{0}$ and $r=r_{0}+\cos \theta _{\pm }$.
We have $\theta _{\pm }=\pi /2$ for the real-energy levels, thus the
real-energy levels of the two loops are always located on the surface ($%
r=r_{0}$) of the $\varphi $-$k$ torus, while the complex energy levels of
the two loops are shifted to outside $(r>r_{0})$ or inside $(r<r_{0})$ the
surface of the $\varphi $-$k$ torus. The graphic eigenstates and energy
bands of the non-Hermitian SSH ladder in the broken parity-time symmetric
phase are depicted in Supplemental Figures~\ref{fig1}(a)-(d) as a comparison
with those in the real-energy gapless phase arising from exceptional points
shown in Figs.~1(e)-1(h) of the Letter.
\clearpage
\end{widetext}

\begin{thebibliography}{999}
\bibitem{Hasan} M. Z. Hasan and C. L. Kane, Colloquium: Topological
insulators, Rev. Mod. Phys. \textbf{82}, 3045 (2010).

\bibitem{XLQ} X. L. Qi and S. C. Zhang, Topological insulators and
superconductors, Rev. Mod. Phys. \textbf{83}, 1057 (2011).

\bibitem{CKC} C. K. Chiu, J. C. Y. Teo, A. P. Schnyder, and S. Ryu,
Classification of topological quantum matter with symmetries, Rev. Mod.
Phys. \textbf{88}, 035005 (2016).

\bibitem{HMW} H. M. Weng, R. Yu, X. Hu, X. Dai, and Z. Fang, Quantum
anomalous Hall effect and related topological electronic states, Adv. Phys.
\textbf{64}, 227 (2015).

\bibitem{LUL} L. Lu, J. D. Joannopoulos, and M. Solja\v{c}i\'{c},
Topological photonics, Nat. Photon. \textbf{8}, 821 (2014); Topological
states in photonic systems, Nat. Phys. \textbf{12}, 626 (2016).

\bibitem{Khanikaev} A. B. Khanikaev and G. Shvets, Two-dimensional
topological photonics, Nat. Photon. \textbf{11}, 763 (2017).

\bibitem{Goldman} N. Goldman, J. C. Budich, and P. Zoller, Topological
quantum matter with ultracold gases in optical lattices, Nat. Phys. \textbf{%
12}, 639 (2016).

\bibitem{Cooper} N. R. Cooper, J. Dalibard, and I. B. Spielman, Topological
bands for ultracold atoms, Rev. Mod. Phys. \textbf{91}, 015005 (2019).

\bibitem{OzawaT} T. Ozawa, H. M. Price, A. Amo, N. Goldman, M. Hafezi, L.
Lu, M. Rechtsman, D. Schuster, J. Simon, O. Zilberberg, and I. Carusotto,
Topological photonics, Rev. Mod. Phys. \textbf{91}, 015006 (2019).

\bibitem{Hafezi} M. Hafezi, E. A. Demler, M. D. Lukin, and J. M. Taylor,
Robust optical delay lines with topological protection, Nat. Phys. \textbf{7}%
, 907 (2011); M. Hafezi, S. Mittal, J. Fan, A. Migdall, and J. M. Taylor,
Imaging topological edge states in silicon photonics, Nat. Photon. \textbf{7}%
, 1001 (2013).

\bibitem{Rechtsman} M. C. Rechtsman, J. M. Zeuner, Y. Plotnik, Y. Lumer, D.
Podolsky, F. Dreisow, S. Nolte, M. Segev, and A. Szameit, Photonic Floquet
topological insulators, Nature (London) \textbf{496}, 196 (2013).

\bibitem{WJChen} W.-J. Chen, S.-J. Jiang, X.-D. Chen, B. Zhu, L. Zhou, J.-W.
Dong, and C. T. Chan, Experimental realization of photonic topological
insulator in a uniaxial metacrystal waveguide, Nat. Commun. \textbf{5}, 5782
(2014).

\bibitem{Mukherjee} S. Mukherjee, A. Spracklen, M. Valiente, E. Andersson,
P. \"{O}hberg, N. Goldman, and R. R. Thomson, Experimental observation of
anomalous topological edge modes in a slowly driven photonic lattice, Nat.
Commun. \textbf{8}, 13918 (2017).

\bibitem{Bandres} M. A. Bandres, M. C. Rechtsman, and M. Segev, Topological
Photonic Quasicrystals: Fractal Topological Spectrum and Protected
Transport, Phys. Rev. X \textbf{6}, 011016 (2016).

\bibitem{Lin} Q. Lin, M. Xiao, L. Yuan, and S. Fan, Photonic Weyl point in a
twodimensional resonator lattice with a synthetic frequency dimension, Nat.
Commun. 7, 13731 (2016).

\bibitem{YDChong16} D. Leykam and Y. D. Chong, Edge Solitons in Nonlinear
Photonic Topological Insulators, Phys. Rev. Lett. \textbf{117}, 143901
(2016).

\bibitem{JWDong} J.-W. Dong, X.-D. Chen, H. Zhu, Y. Wang, and X. Zhang,
Valley photonic crystals for control of spin and topology, Nat. Mater.
\textbf{16}, 298 (2017).

\bibitem{Rechtsman17} J. Noh, S. Huang, D. Leykam, Y. D. Chong, K. P. Chen,
M. C. Rechtsman, Experimental observation of optical Weyl points and Fermi
arc-like surface states, Nat. Phys. \textbf{13}, 611 (2017).

\bibitem{Mittal} S. Mittal, J. Fan, S. Faez, A. Migdall, J.\thinspace M.
Taylor, and M. Hafezi, Topologically Robust Transport of Photons in a
Synthetic Gauge Field, Phys. Rev. Lett. \textbf{113}, 087403 (2014); S.
Mittal, S. Ganeshan, J. Fan, A. Vaezi, and M. Hafezi, Measurement of
topological invariants in a 2D photonic system, Nat. Photon. \textbf{10},
180 (2016); S. Mittal, E. A. Goldschmidt, and M. Hafezi, A topological
source of quantum light, Nature \textbf{561}, 502 (2018).

\bibitem{Klembt} S. Klembt, T. H. Harder, O. A. Egorov, K. Winkler, R. Ge,
M. A. Bandres, M. Emmerling, L. Worschech, T. C. H. Liew, M. Segev, C.
Schneider, and S. H\"{o}fling, Exciton-polariton topological insulator,
Nature \textbf{562}, 552 (2018).

\bibitem{LF} L. Fu and C. L. Kane, Superconducting Proximity Effect and
Majorana Fermions at the Surface of a Topological Insulator, Phys. Rev.
Lett. \textbf{100}, 096407 (2008).

\bibitem{RML} R. M. Lutchyn, J. D. Sau, and S. D. Sarma, Majorana Fermions
and a Topological Phase Transition in Semiconductor-Superconductor
Heterostructures, Phys. Rev. Lett. \textbf{105}, 077001 (2010).

\bibitem{VM} V. Mourik, K. Zuo, S. M. Frolov, S. R. Plissard, E. P. A. M.
Bakkers, and L. P. Kouwenhoven, Signatures of Majorana Fermions in Hybrid
Superconductor-Semiconductor Nanowire Devices, Science \textbf{336}, 1003
(2012); S. N. Perge, I. K. Drozdov, J. Li, H. Chen, S. Jeon, J. Seo, A. H.
MacDonald, B. A. Bernevig, and A. Yazdani, Observation of Majorana fermions
in ferromagnetic atomic chains on a superconductor, Science \textbf{346},
602 (2014).

\bibitem{YO} Y. Oreg, G. Refael, and F. von Oppen, Helical Liquids and
Majorana Bound States in Quantum Wires, Phys. Rev. Lett. \textbf{105},
177002 (2010).

\bibitem{AHCN} A. H. C. Neto, F. Guinea, N. M. R. Peres, K. S. Novoselov,
and A. K. Geim, The electronic properties of graphene, Rev. Mod. Phys.
\textbf{81}, 109 (2009).

\bibitem{JAS} J. A. Steinberg, S. M. Young, S. Zaheer, C. L. Kane, E. J.
Mele, and A. M. Rappe, Bulk Dirac Points in Distorted Spinels, Phys. Rev.
Lett. \textbf{112}, 036403 (2014).

\bibitem{JX} J. Xiong, S. K. Kushwaha, T. Liang, J. W. Krizan, M.
Hirschberger, W. Wang, R. J. Cava, and N. P. Ong, Evidence for the chiral
anomaly in the Dirac semimetal Na$_{3}$Bi, Science \textbf{350}, 413 (2015).

\bibitem{ZKL} Z. K. Liu, J. Jiang, B. Zhou, Z. J. Wang, Y. Zhang, H. M.
Weng, D. Prabhakaran, S-K. Mo, H. Peng, P. Dudin, T. Kim, M. Hoesch, Z.
Fang, X. Dai, Z. X. Shen, D. L. Feng, Z. Hussain, and Y. L. Chen, A stable
three-dimensional topological Dirac semimetal Cd$_{3}$As$_{2}$, Nat. Mater.
\textbf{13}, 677 (2014); Z. K. Liu, B. Zhou, Y. Zhang, Z. J. Wang, H. M.
Weng, D. Prabhakaran, S. K. Mo, Z. X. Shen, Z. Fang, X. Dai, Z. Hussain, and
Y. L. Chen, Discovery of a Three-Dimensional Topological Dirac Semimetal, Na$%
_{3}$Bi, Science \textbf{343}, 864 (2014).

\bibitem{SMY} S. M. Young, S. Zaheer, J. C. Y. Teo, C. L. Kane, E. J. Mele,
and A. M. Rappe, Dirac Semimetal in Three Dimensions, Phys. Rev. Lett.
\textbf{108}, 140405 (2012).

\bibitem{ZW} Z. J. Wang, Y. Sun, X. Q. Chen, C. Franchini, G. Xu, H. M.
Weng, X. Dai, and Z. Fang, Dirac semimetal and topological phase transitions
in A$_{3}$Bi (A=Na, K, Rb), Phys. Rev. B \textbf{85}, 195320 (2012).

\bibitem{BQL} B. Q. Lv, N. Xu, H. M. Weng, J. Z. Ma, P. Richard, X. C.
Huang, L. X. Zhao, G. F. Chen, C. E. Matt, F. Bisti, V. N. Strocov, J.
Mesot, Z. Fang, X. Dai, T. Qian, M. Shi, and H. Ding, Observation of Weyl
nodes in TaAs, Nat. Phys. \textbf{11}, 724 (2015).

\bibitem{MH} M. Hirschberger, S. Kushwaha, Z. J. Wang, Q. Gibson, S. H.
Liang, C. A. Belvin, B. A. Bernevig, R. J. Cava, and N. P. Ong, The chiral
anomaly and thermopower of Weyl fermions in the half-Heusler GdPtBi, Nat.
Mater. \textbf{15, }1161 (2016); H. M. Weng, C. Fang, Z. Fang, B. A.
Bernevig, and X. Dai, Weyl Semimetal Phase in Noncentrosymmetric
Transition-Metal Monophosphides, Phys. Rev. X \textbf{5}, 011029 (2015).

\bibitem{SMH} S. M. Huang, S. Y. Xu, I. Belopolski, C. C. Lee, G. Q. Chang,
B. K. Wang, N. Alidoust, G. Bian, M. Neupane, C. Zhang, S. Jia, A. Bansil,
H. Lin, and M. Z. Hasan, A Weyl Fermion semimetal with surface Fermi arcs in
the transition metal monopnictide TaAs class, Nat. Commun. \textbf{6}, 7373
(2015).

\bibitem{SYX} S. Y. Xu, I. Belopolski, N. Alidoust, M. Neupane, G. Bian, C.
L. Zhang, R. Sankar, G. Q. Chang, Z. J. Yuan, C. C. Lee, S. M. Huang, H.
Zheng, J. Ma, D. S. Sanchez, B. K. Wang, A. Bansil, F. C. Chou, P. P.
Shibayev, H. Lin, S. Jia, and M. Z. Hasan, Discovery of a Weyl fermion
semimetal and topological Fermi arcs, Science \textbf{349}, 613 (2015).

\bibitem{Vishwanath} N. P. Armitage, E. J. Mele, and A. Vishwanath, Weyl and
Dirac semimetals in three-dimensional solids, Rev. Mod. Phys. \textbf{90},
015001 (2018).

\bibitem{Yang} K. Y. Yang, Y. M. Lu, and Y. Ran, Quantum Hall effects in a
Weyl semimetal: Possible application in pyrochlore iridates, Phys. Rev. B
\textbf{84,} 075129 (2011).

\bibitem{Burkov} A. A. Burkov and L. Balents, Weyl Semimetal in a
Topological Insulator Multilayer, Phys. Rev. Lett. \textbf{107,} 127205
(2011).

\bibitem{Xu} G. Xu, H. Weng, Z. Wang, X. Dai, and Z. Fang, Chern Semimetal
and the Quantized Anomalous Hall Effect in HgCr$_{2}$Se$_{4}$, Phys. Rev.
Lett. \textbf{107,} 186806 (2011).

\bibitem{Kim} W. W. Krempa and Y. B. Kim, Topological and magnetic phases of
interacting electrons in the pyrochlore iridates, Phys. Rev. B \textbf{85,}
045124 (2012).

\bibitem{Wang1} Z. Wang, H. Weng, Q. Wu, X. Dai, and Z. Fang,
Three-dimensional Dirac semimetal and quantum transport in Cd$_{3}$As$_{2}$,
Phys. Rev. B \textbf{88,} 125427 (2013).

\bibitem{Hou1} J. M. Hou, Hidden-Symmetry-Protected Topological Semimetals
on a Square Lattice, Phys. Rev. Lett. \textbf{111,} 130403 (2013).

\bibitem{Sama} K. Sun, W. V. Liu, A. Hemmerich, and S. D. Sarma, Topological
semimetal in a fermionic optical lattice, Nat. Phys. \textbf{8,} 67 (2012).

\bibitem{Neupane} M. Neupane, S. Y. Xu, R. Sankar, N. Alidoust, G. Bian, C.
Liu, I. Belopolski, T. R. Chang, H. T. Jeng, H. Lin, A. Bansil, F. C. Chou,
and M. Z. Hasan, Observation of a three-dimensional topological Dirac
semimetal phase in high-mobility Cd$_{3}$As$_{2}$, Nat. Commun. \textbf{5},
3786 (2014).

\bibitem{Lu} L. Lu, Z. Y. Wang, D. X. Ye, L. X. Ran, L. Fu, J. D.
Joannopoulos, and M. Solja\v{c}i\'{c}, Experimental observation of Weyl
points, Science \textbf{349}, 622 (2015).

\bibitem{Li} C. Li, S. Lin, G. Zhang, and Z. Song, Topological nodal points
in two coupled Su-Schrieffer-Heeger chains, Phys. Rev. B \textbf{96}, 125418
(2017).

\bibitem{Bender} C. M. Bender and S. Boettcher, Real Spectra in
non-Hermitian Hamiltonians Having $\mathcal{PT}$ Symmetry, Phys. Rev. Lett.
\textbf{80}, 5243 (1998).

\bibitem{Ali1} A. Mostafazadeh, Pseudo-Hermiticity versus $\mathcal{PT}$%
-symmetry: The necessary condition for the reality of the spectrum of a
non-Hermitian Hamiltonian, J. Math. Phys. \textbf{43}, 205 (2002).

\bibitem{NMBook} N. Moiseyev, Non-Hermitian Quantum Mechanics (Cambridge
University Press, Cambridge, UK, 2011).

\bibitem{Ruschhaupt} A. Ruschhaupt, F. Delgado, and J. G. Muga, Physical
realization of $\mathcal{PT}$-symmetric potential scattering in a planar
slab waveguide, J. Phys. A \textbf{38}, L171 (2005).

\bibitem{Klaiman} S. Klaiman, U. Guenther, and N. Moiseyev, Visualization of
Branch Points in $\mathcal{PT}$-Symmetric Waveguides, Phys. Rev. Lett.
\textbf{101}, 080402 (2008).

\bibitem{Makris08} R. El-Ganainy, K. G. Makris, D. N. Christodoulides, and
Z. H. Musslimani, Opt. Lett. 32, 2632 (2007); K. G. Makris, R. El-Ganainy,
D. N. Christodoulides, and Z. H. Musslimani, Beam Dynamics in $\mathcal{PT}$
Symmetric Optical Lattices, Phys. Rev. Lett. \textbf{100}, 103904 (2008); Z.
H. Musslimani, K. G. Makris, R. El-Ganainy, and D. N. Christodoulides,
Optical Solitons in $\mathcal{PT}$ Periodic Potentials, Phys. Rev. Lett.
\textbf{100}, 030402 (2008).

\bibitem{Guo} A. Guo, G. J. Salamo, D. Duchesne, R. Morandotti, M.
Volatier-Ravat, V. Aimez, G. A. Siviloglou, and D. N. Christodoulides,
Observation of $\mathcal{PT}$-Symmetry Breaking in Complex Optical
Potentials, Phys. Rev. Lett. \textbf{103}, 093902 (2009).

\bibitem{Ruter} C. E. R\"{u}ter, K. G. Makris, R. El-Ganainy, D. N.
Christodoulides, M. Segev, and D. Kip, Observation of parity--time symmetry
in optics, Nat. Phys. \textbf{6}, 192 (2010).

\bibitem{Peng} B. Peng, S. K. \"{O}zdemir, F. Lei, F. Monifi, M. Gianfreda,
G. L. Long, S. Fan, F. Nori, C. M. Bender and L. Yang,
Parity--time-symmetric whispering-gallery microcavities, Nat. Phys. \textbf{%
10}, 394 (2014).

\bibitem{Chang} L. Chang, X. Jiang, S. Hua, C. Yang, J. Wen, L. Jiang, G.
Li, G. Wang and M. Xiao, Parity--time symmetry and variable optical
isolation in active--passive-coupled microresonators, Nat. Photon. \textbf{8}%
, 524 (2014).

\bibitem{XZhang} L. Feng, Z. J. Wong, R. M. Ma, Y. Wang, and X. Zhang,
Single-mode laser by parity-time symmetry breaking, Science \textbf{346},
972 (2014).

\bibitem{Fleury} R. Fleury, D. Sounas, and A. Al\`{u}, An invisible acoustic
sensor based on parity-time symmetry, Nat. Commun. \textbf{6}, 5905 (2015).

\bibitem{Makris15} K. G. Makris, Z. H. Musslimani, D. N. Christodoulides,
and S. Rotter, Constant-intensity waves and their modulation instability in
non-Hermitian potentials, Nat. Commun. \textbf{6}, 7257 (2015).

\bibitem{LFengReview} L. Feng, R. El-Ganainy, and L. Ge, Non-Hermitian
photonics based on parity--time symmetry, Nat. Photo. \textbf{11}, 752
(2017).

\bibitem{El} R. El-Ganainy, K. G. Makris, M. Khajavikhan, Z. H. Musslimani,
S. Rotter and D. N. Christodoulides, Non-Hermitian physics and $\mathcal{PT}$
symmetry, Nat. Phys. \textbf{14}, 11 (2018).

\bibitem{Gupta} S. K. Gupta, Y. Zou, X. Y. Zhu, M. H. Lu, L. J. Zhang, X. P.
Liu, and Y. F. Chen, Parity-time Symmetry in Non-Hermitian Complex Media,
arXiv:1803.00794.

\bibitem{Dembowski} C. Dembowski, H. D. Graf, H. L. Harney, A. Heine, W. D.
Heiss, H. Rehfeld, and A. Richter, Experimental Observation of the
Topological Structure of Exceptional Points, Phys. Rev. Lett. \textbf{86},
787 (2001); C. Dembowski, B. Dietz, H. D. Graf, H. L. Harney, A. Heine, W.
D. Heiss, and A. Richter, Encircling an exceptional point, Phys. Rev. E
\textbf{69}, 056216 (2004).

\bibitem{EP} W. D. Heiss and H. L. Harney, The chirality of exceptional
points, Eur. Phys. J. D \textbf{17}, 149 (2001); W. D. Heiss, Exceptional
points of non-Hermitian operators,\ J. Phys. A \textbf{37}, 2455 (2004); M.
V. Berry, Physics of nonhermitian degeneracies,\ Czech. J. Phys. \textbf{54}%
, 1039 (2004).

\bibitem{Uzdin} R. Uzdin, A. Mailybaev, and N. Moiseyev, On the
observability and asymmetry of adiabatic state flips generated by
exceptional points, J. Phys. A \textbf{44}, 435302 (2011).

\bibitem{Zhen} B. Zhen, C. W. Hsu, Y. Igarashi, L. Lu, I. Kaminer, A. Pick,
S. L. Chua, J. D. Joannopoulos and M. Solja\v{c}i\'{c}, Spawning rings of
exceptional points out of Dirac cones, Nature \textbf{525}, 354 (2015).

\bibitem{Doppler} J. Doppler, A. A. Mailybaev, J. Bohm, U. Kuhl, A.
Girschik, F. Libisch, T. J. Milburn, P. Rabl, N. Moiseyev and S. Rotter,
Dynamically encircling an exceptional point for asymmetric mode switching,
Nature \textbf{537}, 76 (2016).

\bibitem{Xu1} H. Xu, D. Mason, L. Jiang, and J. G. E. Harris, Topological
energy transfer in an optomechanical system with exceptional points, Nature
\textbf{537}, 80 (2016).

\bibitem{Wiersig} J. Wiersig, Enhancing the Sensitivity of Frequency and
Energy Splitting Detection by Using Exceptional Points: Application to
Microcavity Sensors for Single Particle Detection, Phys. Rev. Lett. \textbf{%
112}, 203901 (2014).

\bibitem{ZPLiu} Z. P. Liu, J. Zhang, S. K. \"{O}zdemir, B. Peng, H. Jing, X.
Y. L%
%TCIMACRO{\U{a8}}%
%BeginExpansion
\"{}%
%EndExpansion
u, C. W. Li, L. Yang, F. Nori, and Y. X. Liu, Phys. Rev. Lett. \textbf{117},
110802 (2016).

\bibitem{Chen} W. Chen, S. K. \"{O}zdemir, G. Zhao, J. Wiersig, and L. Yang,
Exceptional points enhance sensing in an optical microcavity, Nature \textbf{%
548}, 192 (2017).

\bibitem{Hodaei} H. Hodaei, A. U. Hassan, S. Wittek, H. Garcia-Gracia, R.
El-Ganainy, D. N. Christodoulides and M. Khajavikhan, Enhanced sensitivity
at higher-order exceptional points, Nature \textbf{548}, 187 (2017).

\bibitem{KDing} K. Ding, G. Ma, M. Xiao, Z. Q. Zhang, and C. T. Chan,
Emergence, Coalescence, and Topological Properties of Multiple Exceptional
Points and Their Experimental Realization, Phys. Rev. X \textbf{6}, 021007
(2016).

\bibitem{KDingPRL} K. Ding, G. Ma, Z. Q. Zhang, and C. T. Chan, Experimental
Demonstration of an Anisotropic Exceptional Point, Phys. Rev. Lett. \textbf{%
121}, 085702 (2018).

\bibitem{Zhang} X. L. Zhang, S. B. Wang, B. Hou, and C. T. Chan, Dynamically
Encircling Exceptional Points: In situ Control of Encircling Loops and the
Role of the Starting Point, Phys. Rev. X \textbf{8}, 021066 (2018).

\bibitem{QZhong} Q. Zhong, D. N. Christodoulides, M. Khajavikhan, K. G.
Makris, and R. El-Ganainy, Power-law scaling of extreme dynamics near
higher-order exceptional points, Phys. Rev. A \textbf{97}, 020105(R) (2018).

\bibitem{Midya} B. Midya, H. Zhao, and Feng, L. Non-Hermitian photonics
promises exceptional topology of light, Nat. Commun. \textbf{9}, 2674 (2018).

\bibitem{Miri} M. A. Miri and A. Al\`{u}, Exceptional points in optics and
photonics, Science \textbf{363}, eaar7709 (2019).

\bibitem{Levitov} M. S. Rudner and L. S. Levitov, Topological transition in
a non-hermitian quantum walk, Phys. Rev. Lett. \textbf{102}, 065703 (2009);
M. S. Rudner,M. Levin, and L. S. Levitov, Survival, decay, and topological
protection in non-Hermitian quantum transport, arXiv:1605.07652.

\bibitem{Diehl} S. Diehl, E. Rico, M. A. Baranov, and P. Zoller, Topology by
dissipation in atomic quantum wires, Nat. Phys. \textbf{7}, 971 (2011).

\bibitem{2011} Y. C. Hu and T. L. Hughes, Absence of topological insulator
phases in non-Hermitian $\mathcal{PT}$-symmetric Hamiltonians, Phys. Rev. B
\textbf{84}, 153101 (2011); K. Esaki, M. Sato, K. Hasebe, and M. Kohmoto,
Edge states and topological phases in non-Hermitian systems, Phys. Rev. B
\textbf{84}, 205128 (2011).

\bibitem{Schomerus} H. Schomerus, Topologically protected midgap states in
complex photonic lattices, Opt. Lett. \textbf{38}, 1912 (2013).

\bibitem{Zeuner} J. M. Zeuner, M. C. Rechtsman, Y. Plotnik, Y. Lumer, S.
Nolte, M. S. Rudner, M. Segev, and A. Szameit, Observation of a topological
transition in the bulk of a non-Hermitian system, Phys. Rev. Lett. \textbf{%
115}, 040402 (2015).

\bibitem{YFChen} C. He, X. C. Sun, X. P. Liu, M. H. Lu, Y. Chen, L. Feng,
and Y. F. Chen, Photonic topological insulator with broken time-reversal
symmetry, Proc. Natl. Acad. Sci. U.S.A. \textbf{113}, 4924 (2016).

\bibitem{WZhu} W. Zhu, X. Fang, D. Li, Y. Sun, Y. Li, Y. Jing, and H. Chen,
Simultaneous observation of a topological edge state and exceptional point
in an open and non-Hermitian acoustic system, Phys. Rev. Lett. \textbf{121},
124501 (2018).

\bibitem{HZhao} H. Zhao, S. Longhi, and L. Feng, Robust light state by
quantum phase transition in non-Hermitian optical materials, Sci. Rep.
\textbf{5}, 17022 (2015); B. Midya and L. Feng, Topological multiband
photonic superlattices, Phys. Rev. A \textbf{98}, 043838 (2018).

\bibitem{GQLiang} G. Q. Liang and Y. D. Chong, Optical resonator analog of a
two-dimensional topological insulator, Phys. Rev. Lett. \textbf{110}, 203904
(2013).

\bibitem{Pump} W. Hu, H. Wang, P. P. Shum, and Y. D. Chong, Exceptional
points in a non-Hermitian topological pump, Phys. Rev. B \textbf{95}, 184306
(2017); Y. Chen and H. Zhai, Hall Conductance of a Non-Hermitian Chern
Insulator, Phys. Rev. B \textbf{98}, 245130 (2018); T. Rakovszky, J. K. Asb%
\'{o}th, and A. Alberti, Detecting topological invariants in chiral
symmetric insulators via losses, Phys. Rev. B \textbf{95}, 201407(R) (2017);
T. M. Philip, M. R. Hirsbrunner, and M. J. Gilbert, Loss of Hall
conductivity quantization in a non-Hermitian quantum anomalous Hall
insulator, Phys. Rev. B \textbf{98}, 155430 (2018).

\bibitem{SChen} C. Yin, H. Jiang, L. Li, R. L\"{u}, and S. Chen, Geometrical
meaning of winding number and its characterization of topological phases in
one-dimensional chiral non-Hermitian systems, Phys. Rev. A \textbf{97},
052115 (2018); H. Jiang, C. Yang, and S. Chen, Topological invariants and
phase diagrams for one-dimensional two-band non-Hermitian systems without
chiral symmetry, Phys. Rev. A \textbf{98}, 052116 (2018); H. Jiang, L. J.
Lang, C. Yang, S. L. Zhu, and S. Chen, Interplay of non-Hermitian skin
effects and Anderson localization in non-reciprocal quasiperiodic lattices,
arXiv:1901.09399.

\bibitem{JGong15} J. Gong and Q. H. Wang, Stabilizing non-Hermitian systems
by periodic driving, Phys. Rev. A \textbf{91}, 042135 (2015); L. Zhou, Q. H.
Wang, H. Wang, and J. Gong, Dynamical quantum phase transitions in
non-Hermitian lattices, Phys. Rev. A \textbf{98}, 022129 (2018).

\bibitem{Song} R.Wang, X. Z. Zhang, and Z. Song, Dynamical topological
invariant for non-Hermitian Rice-Mele model, Phys. Rev. A \textbf{98},
042120 (2018); X. Z. Zhang and Z. Song, Partial topological Zak phase and
dynamical confinement in a non-Hermitian bipartite system, Phys. Rev. A
\textbf{99}, 012113 (2019).

\bibitem{HShen2} H. Shen and L. Fu, Quantum Oscillation from in-gap states
and a non-Hermitian Landau level problem, Phys. Rev. Lett. \textbf{121},
026403 (2018).

\bibitem{Takata} K. Takata and M. Notomi, Photonic topological insulating
phase induced solely by gain and loss, Phys. Rev. Lett. \textbf{121}, 213902
(2018).

\bibitem{YXu2} Q. B. Zeng, Y. B. Yang, and Y. Xu, Topological Non-Hermitian
Quasicrystals, arXiv:1901.08060.

\bibitem{Yoko} T. Yoshida, R. Peters, and N. Kawakami, Non-Hermitian
perspective of the band structure in heavy fermion systems, Phys. Rev. B
\textbf{98}, 035141 (2018); K. Yokomizo and S. Murakami, Bloch Band Theory
for Non-Hermitian Systems, arXiv:1902.10958; Z. Y. Ge, Y. R. Zhang, T. Liu,
S. W. Li, H. Fan, and F. Nori, Topological band theory for non-Hermitian
systems from a quantum field viewpoint, arXiv:1903.09985.

\bibitem{Yuce} C. Yuce, Topological phase in a non-Hermitian $\mathcal{PT}$
symmetric system, Phys. Lett. A \textbf{379}, 1213 (2015); Majorana edge
modes with gain and loss, Phys. Rev. A \textbf{93}, 062130 (2016); Edge
states at the interface of non-Hermitian systems, Phys. Rev. A \textbf{97},
042118 (2018); Stable topological edge states in a non-Hermitian four-band
model, Phys. Rev. A \textbf{98}, 012111 (2018).

\bibitem{Lieu} S. Lieu, Topological phases in the non-Hermitian
Su-Schrieffer-Heeger model, Phys. Rev. B \textbf{97}, 045106 (2018);
Topological symmetry classes for non-Hermitian models and connections to the
bosonic Bogoliubov--de Gennes equation, Phys. Rev. B \textbf{98}, 115135
(2018).

\bibitem{Hirschmann} H. Menke and M. M. Hirschmann, Topological quantum
wires with balanced gain and loss, Phys. Rev. B \textbf{95}, 174506 (2017).

\bibitem{Bergholtz} J. Carlstr\"{o}m and E. J. Bergholtz, Exceptional links
and twisted Fermi ribbons in non-Hermitian systems, Phys. Rev. A \textbf{98}%
, 042114 (2018); J. Carlstr\"{o}m, M. Stalhammar, J. C. Budich, and E. J.
Bergholtz, Knotted Non-Hermitian Metals, Phys. Rev. B \textbf{99}, 161115(R)
(2019).

\bibitem{KK} K. Kawabata, K. Shiozaki, and M. Ueda, Anomalous helical edge
states in a non-Hermitian Chern insulator, Phys. Rev. B \textbf{98}, 165148
(2018).

\bibitem{JHou} J. Hou, Z. Li, X. W. Luo, Q. Gu, and C. Zhang, Topological
bands and triply-degenerate points in non-Hermitian hyperbolic
metamaterials, arXiv:1808.06972.

\bibitem{CYZhao} B. X. Wang and C. Y. Zhao, Topological phonon polaritons in
one-dimensional non-Hermitian silicon carbide nanoparticle chains, Phy. Rev.
B \textbf{98}, 165435 (2018); Topological photonic states in one-dimensional
dimerized ultracold atomic chains, Phy. Rev. A \textbf{98}, 023808 (2018).

\bibitem{VMMA} A. A. Zyuzin and A. Yu. Zyuzin, Flat band in disorderdriven
non-Hermitian Weyl semimetals, Phys. Rev. B \textbf{97}, 041203(R) (2018);
K. Moors, A. A. Zyuzin, A. Y. Zyuzin, R. P. Tiwari, and T. L. Schmidt,
Disorder-driven exceptional lines and Fermi ribbons in tilted nodal-line
semimetals, Phy. Rev. B \textbf{99}, 041116(R) (2019); V. M. Martinez
Alvarez and M. D. Coutinho-Filho, Edge states in trimer lattices, Phys. Rev.
A \textbf{99}, 013833 (2019).

\bibitem{KunstTM} F. K. Kunst and V. Dwivedi, Non-Hermitian systems and
topology: A transfer matrix perspective, arXiv:1812.02186; F. K. Kunst, G.
van Miert, and E. J. Bergholtz, Extended Bloch theorem for topological
lattice models with open boundaries, arXiv:1812.03099.

\bibitem{Herviou} L. Herviou, J. H. Bardarson, and N. Regnault, Impact of
Ground Truth Annotation Quality on Performance of Semantic Image
Segmentation of Traffic Conditions, arXiv:1901.00001; H. G. Zirnstein, G.
Refael, and B. Rosenow, Bulk-boundary correspondence for non-Hermitian
Hamiltonians via Green functions, arXiv:1901.11241; M. R. Hirsbrunner, T. M.
Philip, and M. J. Gilbert, Topology and Observables of the Non-Hermitian
Chern Insulator, arXiv:1901.09961.

\bibitem{Schnyder} W. B. Rui, Y. X. Zhao, and A. P. Schnyder, Classification
of massive Dirac models with generic non-Hermitian perturbations, arXiv:
1902.06617.

\bibitem{DasReview} A. Ghatak and T. Das, New topological invariants in
non-Hermitian systems, J. Phys.: Condens. Matter \textbf{31}, 263001 (2019).

\bibitem{TOhashi} T. Ohashi, S. Kobayashi, and Y. Kawaguchi, Generalized
Berry phase for a bosonic Bogoliubov system with exceptional points,
arXiv:1904.08724.

\bibitem{Malzard2} S. Malzard, C. Poli, and H. Schomerus, Topologically
Protected Defect States in Open Photonic Systems with Non-Hermitian
Charge-Conjugation and Parity-Time Symmetry, Phys. Rev. Lett. \textbf{115},
200402 (2015).

\bibitem{JL} L. Jin, Topological phases and edge states in a non-Hermitian
trimerized optical lattice, Phys. Rev. A \textbf{96}, 032103 (2017); L. Jin,
P. Wang, and Z. Song, Su-Schrieffer-Heeger chain with one pair of $\mathcal{%
PT}$-symmetric defects, Sci. Rep. \textbf{7}, 5903 (2017).

\bibitem{LS} S. Lin and Z. Song, Wide-range-tunable Dirac-cone band
structure in a chiral-time-symmetric non-Hermitian system, Phys. Rev. A
\textbf{96}, 052121 (2017).

\bibitem{PXue} L. Xiao, X. Zhan, Z. H. Bian, K. K. Wang, X. Zhang, X. P.
Wang, J. Li, K. Mochizuki, D. Kim, N. Kawakami, W. Yi, H. Obuse, B. C.
Sanders, and P. Xue, Observation of topological edge states in
parity-time-symmetric quantum walks, Nat. Phys. \textbf{13}, 1117 (2017).

\bibitem{XNi} K. Kawabata, Y. Ashida, H. Katsura, and M. Ueda,
Parity-time-symmetric topological superconductor, Phys. Rev. B \textbf{98},
085116 (2018); X. Ni, D. Smirnova, A. Poddubny, D. Leykam, Y. Chong, and A.
B. Khanikaev, $\mathcal{PT}$ phase transitions of edge states at $\mathcal{PT%
}$ symmetric interfaces in non-Hermitian topological insulators, Phys. Rev.
B \textbf{98}, 165129 (2018); A. Ghatak and T. Das, Theory of
superconductivity with non-Hermitian and parity-time reversal symmetric
Cooper pairing symmetry, Phys. Rev. B \textbf{97}, 014512 (2018).

\bibitem{LJLPRB} L. J. Lang, Y. Wang, H. Wang, and Y. D. Chong, Effects of
non-Hermiticity on Su-Schrieffer-Heeger defect states, Phys. Rev. B \textbf{%
98}, 094307 (2018).

\bibitem{Koutserimpas} T. T. Koutserimpas, A. Al\`{u}, and R. Fleury,
Parametric amplification and bidirectional invisibility in $\mathcal{PT}$
symmetric time-Floquet systems, Phys. Rev. A \textbf{97}, 013839 (2018).

\bibitem{Cancellieri} E. Cancellieri and H. Schomerus, PC-symmetry protected
edge states in interacting driven-dissipative bosonic systems, Phys. Rev. A
\textbf{99}, 033801 (2019); J. Hou, Z. Li, Q. Gu, and C. Zhang,
Non-Hermitian Photonics based on Charge-Parity Symmetry, arXiv: 1904.05260.

\bibitem{AYoshida} A. Yoshida, Y. Ohtaki, R. Ohtaki, and T. Fukui, Edge
states, corner states, and flat bands in a two-dimensional $\mathcal{PT}$
symmetric system, arXiv:1904.05007.

\bibitem{Xu2} Y. Xu, S. T. Wang, and L. M. Duan, Weyl exceptional rings in a
three-dimensional dissipative cold atomic gas, Phys. Rev. Lett. \textbf{118}%
, 045701 (2017).

\bibitem{Leykam} D. Leykam, K. Y. Bliokh, C. Huang, Y. D. Chong, and F.
Nori, Edge modes, degeneracies, and topological numbers in non-Hermitian
systems, Phys. Rev. Lett. \textbf{118}, 040401 (2017).

\bibitem{Shen} H. Shen, B. Zhen, and L. Fu, Topological band theory for
non-Hermitian Hamiltonians, Phys. Rev. Lett. \textbf{120}, 146402 (2018).

\bibitem{Poli} C. Poli, M. Bellec, U. Kuhl, F. Mortessagne, and H.
Schomerus, Selective enhancement of topologically induced interface states
in a dielectric resonator chain, Nat. Commun. \textbf{6}, 6710 (2015).

\bibitem{Weimann} S. Weimann, M. Kremer, Y. Plotnik, Y. Lumer, S. Nolte, K.
G. Makris, M. Segev, M. C. Rechtsman, and A. Szameit, Topologically
protected bound states in photonic parity--time-symmetric crystals, Nat.
Mater. \textbf{16}, 433 (2017).

\bibitem{MPan} M. Pan, H. Zhao, P. Miao, S. Longhi, and L. Feng, Photonic
zero mode in a non-Hermitian photonic lattice, Nat. Commun. \textbf{9}, 1308
(2018).

\bibitem{Jean} P. St-Jean, V. Goblot, E. Galopin, A. Lemaitre, T. Ozawa, L.
Le Gratiet, I. Sagnes, J. Bloch, and A. Amo, Lasing in topological edge
states of a one-dimensional lattice, Nat. Photon \textbf{11}, 651 (2017).

\bibitem{Parto} M. Parto, S. Wittek, H. Hodaei, G. Harari, M. A. Bandres, J.
Ren, M. C. Rechtsman, M. Segev, D. N. Christodoulides, and M. Khajavikhan,
Edge-mode lasing in 1D topological active arrays, Phys. Rev. Lett. \textbf{%
120}, 113901 (2018).

\bibitem{Feng} H. Zhao, P. Miao, M. H. Teimourpour, S. Malzard, R.
El-Ganainy, H. Schomerus, and L. Feng, Topological hybrid silicon
microlasers, Nat. Commun. \textbf{9}, 981 (2018).

\bibitem{Harari} G. Harari, M. A. Bandres, Y. Lumer, M. C. Rechtsman, Y. D.
Chong, M. Khajavikhan, D. N. Christodoulides, and M. Segev, Topological
insulator laser: Theory, Science \textbf{359}, eaar4003 (2018); M. A.
Bandres, S. Wittek, G. Harari, M. Parto, J. Ren, M. Segev, D.
Christodoulides, and M. Khajavikhan, Topological insulator laser:
Experiments, Science \textbf{359}, eaar4005 (2018).

\bibitem{Kartashov} Y. V. Kartashov and D. V. Skryabin, Two-Dimensional
Topological Polariton Laser, Phys. Rev. Lett. \textbf{122}, 083902 (2019).

\bibitem{Secli} M. Secl\`{\i}, M. Capone, and I. Carusotto, Theory of chiral
edge state lasing in a two-dimensional topological system, arXiv:1901.01290.

\bibitem{TLiu} T. Liu, Y. R. Zhang, Q. Ai, Z. Gong, K. Kawabata, M. Ueda,
and F. Nori, Second-order topological phases in non-Hermitian systems, Phys.
Rev. Lett. \textbf{122}, 076801 (2018).

\bibitem{EzawaPRB} M. Ezawa, Non-Hermitian boundary and interface states in
nonreciprocal higher-order topological metals and electrical circuits, Phys.
Rev. B \textbf{99}, 121411(R) (2019); Electric-circuit realization of
non-Hermitian higher-order topological systems, arXiv:1810.04527.

\bibitem{CHLeeHighOrder} C. H. Lee, L. Li, and J. Gong, Hybrid higher-order
skin-topological modes in non-reciprocal systems, arXiv:1810.11824.

\bibitem{XWLuo} X. W. Luo and C. Zhang, Higher-order topological corner
states induced by gain and loss, arXiv:1903.02448.

\bibitem{Edvardsson} E. Edvardsson, F. K. Kunst, and E. J. Bergholtz,
Non-Hermitian extensions of higher-order topological phases and their
biorthogonal bulk-boundary correspondence, Phys. Rev. B \textbf{99},
081302(R) (2019).

\bibitem{TELee} T. E. Lee, Anomalous edge state in a non-Hermitian lattice,
Phys. Rev. Lett. \textbf{116}, 133903 (2016); Y. Xiong, Why does bulk
boundary correspondence fail in some non-hermitian topological models, J.
Phys. Commun. \textbf{2}, 035043 (2018).

\bibitem{Torres} V. M. Martinez Alvarez, J. E. Barrios Vargas, and L. E. F.
Foa Torres, Non-Hermitian robust edge states in one dimension: Anomalous
localization and eigenspace condensation at exceptional points, Phys. Rev. B
\textbf{97}, 121401(R) (2018); V. M. Martinez Alvarez, J. E. Barrios Vargas,
M. Berdakin, and L. E. F. Foa Torres, Topological states of non-Hermitian
systems, Eur. Phys. J. Spec. Top. \textbf{227}, 1295 (2018).

\bibitem{ZWang} S. Yao and Z. Wang, Edge states and topological invariants
of non-Hermitian systems, Phys. Rev. Lett. \textbf{121}, 086803 (2018); S.
Yao, F. Song, and Z. Wang, Non-hermitian Chern bands, Phys. Rev. Lett.
\textbf{121}, 136802 (2018); F. K. Kunst, E. Edvardsson, J. C. Budich, and
E. J. Bergholtz, Biorthogonal bulk-boundary correspondence in non-Hermitian
systems, Phys. Rev. Lett. \textbf{121}, 026808 (2018).

\bibitem{LJinPRB} L. Jin and Z. Song, Bulk-boundary correspondence in a
non-Hermitian system in one dimension with chiral inversion symmetry, Phys.
Rev. B \textbf{99}, 081103(R) (2019).

\bibitem{JHu} Z. Yang and J. Hu, Non-Hermitian Hopf-link exceptional line
semimetals, Phys. Rev. B \textbf{99}, 081102(R) (2019).

\bibitem{HZhang} H. Wang, J. Ruan, and H. Zhang, Non-Hermitian nodal-line
semimetals with an anomalous bulk-boundary correspondence, Phys. Rev. B
\textbf{99}, 075130 (2019).

\bibitem{CHLee} C. H. Lee and R. Thomale, Anatomy of skin modes and topology
in non-Hermitian systems, arXiv:1809.02125; C. H. Lee, G. Li, Y. Liu, T.
Tai, R. Thomale, and X. Zhang, Tidal surface states as fingerprints of
non-Hermitian nodal knot metals, arXiv:1812.02011.

\bibitem{1819} K. Luo, J. Feng, Y. X. Zhao, and R. Yu, Nodal Manifolds
Bounded by Exceptional Points on Non-Hermitian Honeycomb Lattices and
Electrical-Circuit Realizations, arXiv:1810.09231; D. S. Borgnia, A. J.
Kruchkov, and R. J. Slager, Non-Hermitian Boundary Modes, arXiv:1902.07217;
T.-S. Deng and W. Yi, Non-Bloch topological invariants in a non-Hermtian
domain-wall system, arXiv: 1903.03811; F. Song, S. Yao, and Z. Wang,
Non-Hermitian skin effect and chiral damping in open quantum systems, arXiv:
1904.08432.

\bibitem{ZGong} Z. P. Gong, Y. Ashida, K. Kawabata, K. Takasan, S.
Higashikawa, and M. Ueda, Topological phases of non-Hermitian systems, Phys.
Rev. X \textbf{8}, 031079 (2018).

\bibitem{Kawabata1} K. Kawabata, S. Higashikawa, Z. Gong, Y. Ashida, and M.
Ueda, Topological unification of time-reversal and particle-hole symmetries
in non-Hermitian physics, Nat. Commun. \textbf{10}, 297 (2019).

\bibitem{HZhou} H. Zhou and J. Y. Lee, Periodic Table for Topological Bands
with Non-Hermitian Bernard-LeClair Symmetries, arXiv:1812.10490.

\bibitem{Kawabata2} K. Kawabata, K. Shiozaki, M. Ueda, and M. Sato, Symmetry
and Topology in Non-Hermitian Physics, arXiv:1812.09133.

\bibitem{KawabataEP} K. Kawabata, T. Bessho, and M. Sato, Non-Hermitian
Topology of Exceptional Points, arXiv:1902.08479.

\bibitem{Szameit} A. Szameit, M. C. Rechtsman, O. Bahat-Treidel, and M.
Segev, $\mathcal{PT}$-symmetry in honeycomb photonic lattices, Phys. Rev. A
\textbf{84}, 021806(R) (2011).

\bibitem{Zhou2} H. Zhou, C. Peng, Y. Yoon, C. W. Hsu, K. A. Nelson, L. Fu,
J. D. Joannopoulos, M. Solja\v{c}i\'{c}, and B. Zhen, Observation of bulk
Fermi arc and polarization half charge from paired exceptional points,
Science \textbf{359}, 1009 (2018).

\bibitem{SLin} S. Lin, L. Jin, and Z. Song, Symmetry protected topological
phases characterized by isolated exceptional points, Phys. Rev. B \textbf{99}%
, 165148 (2019).

\bibitem{Kozii} V. Kozii and L. Fu, Non-Hermitian Topological Theory of
Finite-Lifetime Quasiparticles: Prediction of Bulk Fermi Arc Due to
Exceptional Point, arXiv:1708.05841.

\bibitem{Schomerus18} S. Malzard and H. Schomerus, Bulk and edge-state arcs
in non-Hermitian coupled-resonator arrays, Phys. Rev. A \textbf{98}, 033807
(2018).

\bibitem{Budich} J. C. Budich, J. Carlstr\"{o}m, F. K. Kunst, and E. J.
Bergholtz, Symmetry-protected nodal phases in non-Hermitian systems, Phys.
Rev. B \textbf{99}, 041406(R) (2019).

\bibitem{Okugawa} R. Okugawa and T. Yokoyama, Topological exceptional
surfaces in non-Hermitian systems with parity-time and parity-particle-hole
symmetries, Phys. Rev. B \textbf{99}, 041202(R) (2019).

\bibitem{Cerjan1} A. Cerjan, M. Xiao, L. Yuan, and S. Fan, Effects of
non-Hermitian perturbations on Weyl Hamiltonians with arbitrary topological
charges, Phys. Rev. B \textbf{97}, 075128 (2018).

\bibitem{ZhouH} H. Zhou, J. Y. Lee, S. Liu, and B. Zhen, Exceptional
surfaces in $\mathcal{PT}$-symmetric non-Hermitian photonic systems, Optica
\textbf{6}, 190 (2019).

\bibitem{Cerjan2} A. Cerjan, S. Huang, K. P. Chen, Y. Chong, and M. C.
Rechtsman, Experimental realization of a Weyl exceptional ring,
arXiv:1808.09541.

\bibitem{Yoshida} T. Yoshida, R. Peters, N. Kawakami, and Y. Hatsugai,
Symmetry-protected exceptional rings in two-dimensional correlated systems
with chiral symmetry, Phys. Rev. B \textbf{99}, 121101(R) (2019).

\bibitem{RAMolina} J. Gonz\'{a}lez and R. A. Molina, Topological protection
from exceptional points in Weyl and nodal-line semimetals, Phys. Rev. B
\textbf{96}, 045437 (2017); R. A. Molina and J. Gonz\'{a}lez, Surface and 3D
quantum Hall effects from engineering of exceptional points in nodal-line
semimetals, Phys. Rev. Lett. \textbf{120}, 146601 (2018).

\bibitem{Carlstrom} J. Carlstr\"{o}m, M. St\aa lhammar, J. C. Budich, and E.
J. Bergholtz, Knotted non-Hermitian metals, Phys. Rev. B \textbf{99},
161115(R) (2019).

\bibitem{SI} Supplemental Materials for the vector field and the graphic
eigenstates in the broken $\mathcal{PT}$ phase.

\bibitem{Leykam16} D. Leykam, M.\thinspace C. Rechtsman, and Y.\thinspace D.
Chong, Anomalous Topological Phases and Unpaired Dirac Cones in Photonic
Floquet Topological Insulators, Phys. Rev. Lett. \textbf{117}, 013902 (2016).

\bibitem{Xiao} M. Xiao, W. J. Chen, W. Y. He, and C. T. Chan, Synthetic
gauge flux and Weyl points in acoustic systems, Nat. Phys. \textbf{11}, 920
(2015); Z. Yang, F. Gao, X. Shi, X. Lin, Z. Gao, Y. Chong, and B. Zhang,
Topological Acoustics, Phys. Rev. Lett. \textbf{114}, 114301 (2015).

\bibitem{Fleury2} R. Fleury, A. B. Khanikaev, and A. Al\`{u}, Floquet
topological insulators for sound, Nat. Commun. \textbf{7}, 11744 (2016).

\bibitem{HeC} C. He, X. Ni, H. Ge, X. C. Sun, Y. B. Chen, M. H. Lu, X. P.
Liu and Y. F. Chen, Acoustic topological insulator and robust one-way sound
transport, Nat. Phys. \textbf{12}, 1124 (2016).

\bibitem{ChongNC} Y. Wang, L. J. Lang, C. H. Lee, B. Zhang, and Y. D. Chong,
Topologically enhanced harmonic generation in a nonlinear transmission line
metamaterial, Nat. Commun. \textbf{10}, 1102 (2019).

\bibitem{Lee} C. H. Lee, S. Imhof, C. Berger, F. Bayer, J. Brehm, and L. W.
Molenkamp, Tobias Kiessling and Ronny Thomale Topolectrical Circuits,
Commun. Phys. \textbf{1}, 39 (2018).

\bibitem{Ezawa} M. Ezawa, Electric-circuit realization of Hermitian and
non-Hermitian Majorana edge states, arXiv:1902.03716; Electric circuits for
non-Hermitian Chern insulators, arXiv:1904.03823.
\end{thebibliography}
\end{document}